\documentclass[a4paper,11pt]{article}
\usepackage{amsmath, amssymb, a4wide, bm, graphicx, subfig, cancel}
\usepackage{cite}
\usepackage{hyperref}
\newcommand{\SU}[1]{\mathrm{SU}(#1)}

\newcommand{\Sp}[1]{\mathrm{Sp}(#1)}
\newcommand{\U}[1]{\mathrm{U}(#1)}
\newcommand{\Tr}[2]{\mathop{\mathrm{Tr}}_{#1}\left[#2\right]}
\newcommand{\pdiff}[2]{\frac{\partial#1}{\partial#2}}

\newcommand{\nbrack}[1]{\left(#1\right)}
\newcommand{\sbrack}[1]{\left[#1\right]}

\newcommand{\expect}[1]{\langle#1\rangle}
\newcommand{\sep}[1]{\quad\mbox{#1} \quad}
\newcommand{\brm}[1]{\bm{\mathrm{#1}}}
\newcommand{\cbrm}[1]{\overline{\bm{\mathrm{#1}}}}
\newcommand{\vac}[1]{|{\rm vac}\rangle_{#1}}
\def\widerow{\rule{0pt}{2.5ex}\rule[-1.5ex]{0pt}{0pt}}
\def\be{\begin{equation}}
\def\ee{\end{equation}}
\def\ba{\begin{eqnarray}}
\def\ea{\end{eqnarray}}
\def\cN{{\cal N}}
\def\uno{\mbox{1 \kern-.59em {\rm l}}}

\numberwithin{equation}{section}
\numberwithin{figure}{section}
\numberwithin{table}{section}
\begin{document}

\title{{\normalsize DCPT/09/75\hfill\mbox{}\hfill\mbox{}}\\
\vspace{2.5 cm}
\Large{\textbf{Tree Level Metastability and Gauge Mediation in Baryon Deformed SQCD}}}
\vspace{2.5 cm}
\author{James Barnard\\[3ex]
\small{\em Department of Mathematical Sciences, Durham University, Durham DH1 3LE, UK}\\[1.5ex] 
\small james.barnard@durham.ac.uk\\[1.5ex]}
\date{}
\maketitle
\vspace{2ex}
\begin{abstract}
\noindent
We investigate supersymmetric QCD with gauge group $\SU{2}$ and a baryon deformation to the superpotential.   The existence of an uplifted vacuum at the origin with tree level metastability is demonstrated.  When this model is implemented in a direct gauge mediation scenario we therefore find gaugino masses which are comparable to sfermion masses and parameterised by an effective number of messengers $1/8$.  All deformations are well motivated by appealing to the electric theory and an $R$-symmetry.  This $R$-symmetry is explicitly broken by the same term responsible for supersymmetry breaking.  Moreover, the model does not suffer from the Landau pole problem and we find that it can be described in terms of just two scales: the weak scale and a high scale like the Planck or GUT scale.  The model can be tested by searching for new particles at the TeV scale charged under the visible sector gauge group.
\end{abstract}
\newpage

\section{Introduction}

As a supersymmetry (SUSY) breaking scenario, gauge mediation \cite{Dine:1981za, Dimopoulos:1981au, Dine:1981gu, Nappi:1982hm, AlvarezGaume:1981wy, Dimopoulos:1982gm, Dine:1993yw, Dine:1994vc, Dine:1995ag, Giudice:1998bp} presents many advantages.  Chief among these are its general calculability and its lack of flavour problem.  Calculability arises from the fact that the supersymmetry breaking sector can be fully described without appealing to the underlying, incalculable supergravity theory, whereas the flavour problem is solved because the messengers only interact with gauge field supermultiplets in the visible sector.  Such interactions do not generate problematic flavour changing soft terms.  An even more theoretically appealing scenario is that of direct mediation \cite{Affleck:1984xz, Luty:1998vr, Amariti:2007qu, Abel:2007jx, Haba:2007rj, Essig:2008kz}.  Here, the visible sector gauge group is embedded into a flavour symmetry of the supersymmetry breaking sector.  The messengers are thus generated by the dynamics of the theory so there is no need for an additional messenger sector.

Unfortunately direct gauge mediation suffers from two particular problems.  Many models of direct mediation suffer from the Landau pole problem, where the inclusion of messenger fields (charged under both the SUSY breaking and visible sector gauge groups) pushes the theory towards strong coupling in the ultraviolet\footnote{Although it is possible this is not actually a problem after all \cite{Abel:2008tx, Abel:2009bj}}.  Another common problem is the surprisingly small gaugino mass to sfermion mass ratios generated by these models.  Phenomenological constraints that the gaugino masses are above the weak scale thus force very heavy sfermions, leading to fine tuning.

A way to address the smallness of the gaugino masses was discussed in \cite{Komargodski:2009jf}; it appears that they are connected to global properties of the theory's vacua.  For gaugino masses in a given vacuum to be non-zero at leading order in the SUSY breaking\footnote{The SUSY breaking is parameterised by $F/\expect{\phi}^2$ where $\expect{\phi}$ is the VEV that breaks $R$-symmetry.  This is usually small in models of direct mediation so the leading order contribution is dominant, although in known examples with large SUSY breaking the gaugino mass still seems anomalously small.} there must exist lower energy states elsewhere in the pseudomoduli space {\em at tree level}.  Explicit realisations of this idea can be found in the literature: Refs.~\cite{Giveon:2009yu, Koschade:2009qu, Kutasov:2009kb} consider uplifted vacua\footnote{``Uplifted'' in this context refers to a higher metastable vacuum which appears upon the addition of some deformation to the theory.}, whereas \cite{Abel:2009ze} considers a lower energy vacuum pulled in from infinity.  Metastability occurs at tree level in all of these examples\footnote{Tree level metastability was first discussed in \cite{Dienes:2008gj}, using both $F$ and $D$-terms.  We will only discuss $F$-term SUSY breaking here.} and, as such, they include arbitrary deformations that break $R$-symmetry.  Rather disturbingly this renders the theories non-generic.  Moreover, some degree of fine tuning appears to be necessary.  The question then arises as to whether the uplifted vacuum scenario can be put on a more natural footing within the framework of Intriligator, Seiberg and Shih (ISS) \cite{Intriligator:2006dd}.  In this paper we will see that indeed it can be, using the approach of \cite{Murayama:2006yf} in which all deformations and scales appearing in the IR magnetic theory are motivated by appealing to the UV electric theory.

To achieve this we analyse uplifted vacua in massive supersymmetric QCD (SQCD), as in \cite{Giveon:2009yu}, but restrict our attention to the magnetic gauge group of $\SU{2}$.  In this case the vacua can be stabilised by baryon deformations\footnote{Supersymmetry breaking in baryon deformed SQCD was previously studied in \cite{Abel:2007jx} but the vacua presented here are distinct; they are uplifted with respect to those in \cite{Abel:2007jx} so have significantly different phenomenological properties.}.  Baryon deformations in the magnetic theory map to non-renormalisable operators in the electric theory so come with a natural suppression.  However, the emergent mass hierarchy and the smallness of the magnetic gauge group eliminate the Landau pole problem typically encountered in $\SU{5}$ GUT models of direct mediation.  In addition, it is possible to have only two distinct mass scales in the electric theory: the weak scale $\sim100\mbox{ GeV}$ and some high scale, such as the Planck scale or the GUT scale. Both of these scales have obvious physical significance.  Any extra deformations required to stabilise the vacuum or mediate the SUSY breaking to the visible sector are also well understood.  Turning off the SUSY breaking restores an $R$-symmetry that forbids all operators other than those required.  It is well known that breaking $R$-symmetry is already a necessary requirement for metastable SUSY breaking in generic models \cite{Nelson:1993nf}.  In this model the mass term in the electric theory that induces SUSY breaking (as in the ISS model) actually becomes the order parameter for both the SUSY breaking {\em and} the $R$-symmetry breaking.  Finally, the main example we present makes two concrete predictions.  The first is to fix the effective number of messengers at $1/8$.  The second is the existence of new, TeV scale particles charged under the Standard Model.  Such particles should be clearly visible at the LHC so the theory can readily be tested.

This paper is organised as follows.  In \S\ref{sec:rev} we briefly review the arguments of \cite{Komargodski:2009jf} and show how they were applied to uplifted vacua in massive SQCD by the authors of \cite{Giveon:2009yu}.  In \S\ref{sec:BD} we see how a vacuum at the origin can instead be stabilised by adding baryon deformations to the superpotential when the magnetic gauge group is $\SU{2}$.  In \S\ref{sec:G} and \S\ref{sec:DM} we show how this type of model can be implemented in direct mediation scenarios and discuss the phenomenological ramifications in \S\ref{sec:PV}.  We conclude in \S\ref{sec:conc}.

\section{Review\label{sec:rev}}

\subsection{Gaugino masses and gauge mediation\label{sec:Kom}}

We begin with a brief review of \cite{Komargodski:2009jf}, explaining why large gaugino masses require uplifted vacua.  For the simplest general model of gauge mediation \cite{Cheung:2007es} we have a superpotential
\be
W=\eta^i_j\phi\rho_i\tilde{\rho}^j+m^i_j\rho_i\tilde{\rho}^j\,.
\ee
$\phi$ is a spurion whose $F$-term gets a SUSY breaking VEV and the $\rho$'s are messengers which live in the (conjugate) fundamental representation of the visible sector gauge group, e.g.\ $\brm{5}\oplus\cbrm{5}$ of $\SU{5}$.  $\eta$ and $m$ are coupling constants.  It was shown in \cite{Ray:2006wk} that the scalar component of $\phi$ must be massless at tree level, i.e.\ it is a pseudomodulus or flat direction of the theory.  The gaugino masses are consequently expected to be a function of $\phi$, where we now use $\phi$ to denote the scalar component.  Indeed, one finds \cite{Cheung:2007es}
\be
m_{\lambda}\sim\partial_{\phi}\ln\det\nbrack{\phi\eta^i_j+m^i_j}
\ee
at lowest order.  The gaugino mass can only be non-zero at this order if $\det(\phi\eta^i_j+m^i_j)$, which is a polynomial in $\phi$, is $\phi$ dependent.  If this is the case there must be some value $\phi_0$ for which $\det(\phi_0\eta^i_j+m^i_j)=0$.  Since $\phi\eta^i_j+m^i_j$ is also the mass matrix of fermionic messengers this means there are massless fermionic messengers at $\phi_0$.  In \cite{Komargodski:2009jf} the authors demonstrate that a massless fermionic messenger always leads to a massless or tachyonic scalar.  If one follows the direction of this scalar, and the messenger does not decouple from the theory, the energy can be shown to decrease.  Hence the only way to achieve $m_{\lambda}\neq0$ is for there to exist {\em tree level} states at lower energy somewhere in the pseudomoduli space.

\subsection{The ISS model\label{sec:ISS}}

In order to define notation we will start with a brief discussion of the ISS model \cite{Intriligator:2006dd}.  Consider $\cN=1$ SQCD with gauge group $\SU{N}$ and flavour symmetry $\SU{N_f}\times\SU{N_f}$.  The matter content of the theory consists of quarks $Q$ and antiquarks $\tilde{Q}$ and there is no superpotential.  For $N_f>N$ this theory has a dual description \cite{Seiberg:1994pq, Intriligator:1995au}: SQCD with gauge group $\SU{n}$ where
\be
n=N_f-N\,.
\ee
The global symmetry group is $\SU{N_f}\times\SU{N_f}\times\U{1}_B\times\U{1}_R$ and the matter content consists of quarks $q$, antiquarks $\tilde{q}$ and a gauge singlet meson field $\Phi$.  It is summarised in Table \ref{tab:ISSmm}.
\begin{table}[!htb]
\be
\begin{array}{|c|ccccc|}\hline
\widerow & \SU{n} & \SU{N_f} & \SU{N_f} & \U{1}_B & \U{1}_R \\\hline
\widerow q_{ia} & \brm{n} & \brm{N_f} & \brm{1} & \frac{1}{n} & 1-\frac{n}{N_f} \\
\widerow \tilde{q}^{ja} & \cbrm{n} & \brm{1} & \cbrm{N_f} & -\frac{1}{n} & 1-\frac{n}{N_f} \\
\widerow \Phi^i_j & \brm{1} & \cbrm{N_f} & \brm{N_f} & 0 & \frac{2n}{N_f} \\\hline
\end{array}\nonumber
\ee
\caption{\em The matter content of the magnetic theory.  Indices $i$ and $j$ denote flavour, the index $a$ denotes colour.\label{tab:ISSmm}}
\end{table}
The dual theory has superpotential
\be
\frac{1}{h}W_{\rm mg}=\tilde{q}^{ja}\Phi^i_jq_{ia}
\ee
where $h$ is a coupling constant\footnote{We will assume all superpotential coupling constants are real and positive throughout this paper to simplify the analysis.}.  We typically refer to the $\SU{N}$ theory as the electric theory and the $\SU{n}$ theory as the magnetic theory.  There is a one to one correspondence between gauge invariant operators in the two theories.  The meson map is the obvious choice
\be
\tilde{Q}_j^aQ^i_a\leftrightarrow\Lambda\Phi^i_j
\ee
and the baryon map is
\begin{align}\label{eq:ISSBmap}
\epsilon^{(N)}Q^N & \leftrightarrow\epsilon^{(N_f)}\Lambda^{N-n}\epsilon^{(n)}q^n\nonumber\\
\epsilon_{(N)}\tilde{Q}^N & \leftrightarrow\epsilon_{(N_f)}\Lambda^{N-n}\epsilon_{(n)}\tilde{q}^n
\end{align}
with $\epsilon^{(N)}$ denoting contraction with a rank $N$ alternating tensor.  $\Lambda$ is the scale of the theory\footnote{For simplicity, we assume both electric and magnetic theories have the same scale throughout} which is included to ensure the dimensions of the operators match.

It is well known \cite{Intriligator:2006dd} that this theory has a SUSY breaking vacuum for $N+1\le N_f<\frac{3}{2}N$ (or equivalently $1\le n<\frac{1}{3}N_f$) when the electric theory is deformed by a quark mass term
\be\label{eq:ISSWel}
\frac{1}{h}W_{\rm el}=m_Q\tilde{Q}_i^aQ^i_a
\ee
with $m_Q\ll\Lambda$.  This term breaks the $\SU{N_f}\times\SU{N_f}$ flavour symmetry to a diagonal $\SU{N_f}$ subgroup and also leaves only an anomalous $R$-symmetry, which is broken by non-perturbative effects.  In the magnetic theory the superpotential is deformed to
\be\label{eq:ISSWmg}
\frac{1}{h}W_{\rm mg}=\tilde{q}^{ja}\Phi^i_jq_{ia}-\mu^2\Phi^i_i
\ee
with $\mu^2\sim\Lambda m_Q$, so the $F$-term for the meson field is
\be
F_{\Phi}{}_i^j=h\nbrack{\tilde{q}^{ja}q_{ia}-\mu^2\delta_i^j}\,.
\ee
The operator $\tilde{q}^{ja}q_{ia}$ is at most a rank $n$ matrix, whereas $\delta_i^j$ is rank $N_f$.  Since $n=N_f-N$ we know that $N_f>n$ so the $F_{\Phi}{}_i^j$ cannot all be set to zero.  Supersymmetry is thus broken by the rank condition and the vacuum is
\be\label{eq:ISSvac}
\vac{\rm ISS}:\quad\quad q=\tilde{q}^T=\mu\nbrack{\begin{array}{c}\uno_n\\0\end{array}}\,,\quad\Phi=0\,,\quad V_{\rm tree}=\nbrack{N_f-n}h^2\mu^4
\ee
where $\uno_n$ denotes the $n\times n$ identity matrix.  Some of the components of $\Phi$ are pseudomoduli but are stabilised by the Coleman-Weinberg potential at one-loop.  When non-perturbative effects are taken into account we find a supersymmetric vacuum at large $\Phi$.  Hence this vacuum is only metastable, but remains globally stable at tree level.  This metastability is important but the fact that it appears only after non-perturbative effects are considered precludes the possibility that it allows for anything other than small gaugino masses in a gauge mediation scenario.

\subsection{Uplifted vacua in massive SQCD\label{sec:UpV}}

In \cite{Giveon:2009yu} some other vacua of massive SQCD were investigated.  The vacuum of \eqref{eq:ISSvac} allows the matrix $\tilde{q}q$ to fulfil its maximal rank, $n$, but by reducing the rank one can find other stationary points at higher energies than $\vac{\rm ISS}$.  Specifically, in $\vac{\rm ISS}$ we have
\be
\tilde{q}q=\mu\nbrack{\begin{array}{ll}\uno_n & 0_{n\times(N_f-n)} \\ 0_{(N_f-n)\times n} & 0_{(N_f-n)\times(N_f-n)} \end{array}}
\ee
but we could have
\be\label{eq:UpVsp}
\tilde{q}q=\mu\nbrack{\begin{array}{ll}\uno_{n-k} & 0_{(n-k)\times(N_f+k-n)} \\ 0_{(N_f+k-n)\times(n-k)} & 0_{(N_f+k-n)\times(N_f+k-n)} \end{array}}
\ee
for some positive integer $k\le n$.  Such states remain stationary points of the tree level potential\footnote{Assuming we are on a $D$-flat direction.}
\be
V_{\rm tree}=\sum|F|^2\,.
\ee
but have a higher vacuum energy $\nbrack{N_f+k-n}h^2\mu^4$ and so are uplifted with respect to $\vac{\rm ISS}$.  It was shown in \cite{Giveon:2009yu} that these uplifted vacua can be stabilised in the quantum theory by adding an operator $\tilde{Q}Q\tilde{Q}Q$ with non-trivial index contraction to the electric theory.  In addition, the parameter $\mu$ must be split into two different mass scales, $\mu_1$ and $\mu_2$, by replacing $m_Q$ with $m_1$ and $m_2$ in the electric theory.  As discussed in \S\ref{sec:Kom}, using an uplifted vacuum in a direct mediation scenario allows the visible sector gauginos to acquire a similar mass to the sfermions.  This was shown explicitly to be the case in \cite{Giveon:2009yu}.  Unfortunately that model required both the introduction of an additional low mass scale and fine tuning between $m_1$, $m_2$ and the $\tilde{Q}Q\tilde{Q}Q$ coupling constant.  What's more, the relative sizes of the scales are somewhat mysterious, with large ratios appearing for an unexplained reasons and the Landau pole problem persists in most cases as well.

Suppose we focus on the state $q=\tilde{q}^T=0$ in undeformed massive SQCD; the extremal case of \eqref{eq:UpVsp} with $k=n$.  The VEV of the tree level potential in this state is $N_fh^2\mu^4$.  Unlike the ISS vacuum, {\em all} components of $\Phi$ are massless at tree level so should be considered as pseudomoduli, but we can use the remaining $\SU{N_f}$ flavour symmetry to make $\Phi$ diagonal.  To investigate the stability of the stationary point we calculate the bosonic and fermionic mass squared matrices
\be
m_B^2=\nbrack{\begin{array}{cc}
W^{\dag ac}W_{cb} & W^{\dag abc}W_c \\
W_{abc}W_{\dag c} & W_{ac}W^{\dag cb} \end{array}}
\sep{and}
m_f^2=\nbrack{\begin{array}{cc}
W^{\dag ac}W_{cb} & 0 \\
0 & W_{ac}W^{\dag cb} \end{array}}
\ee
respectively.  Using the superpotential \eqref{eq:ISSWmg}, the eigenstates and corresponding eigenvalues in the scalar quark sector around $q=\tilde{q}^T=0$ are
\be
q\pm\tilde{q}^{\dag} \quad:\quad \hat{m}_B^2=h^2\nbrack{|\Phi|^2\mp\mu^2}
\ee
with equivalent relations for the conjugate states.  We see that $q+\tilde{q}^{\dag}$ becomes tachyonic when $|\Phi|<\mu$, in which case the theory flows back to the ISS vacuum \eqref{eq:ISSvac}.  As a result, we need to stabilise $|\Phi|$ above $\mu$ for the stationary point to be a vacuum of the theory.

This might be accomplished at one-loop by the Coleman-Weinberg potential \cite{Coleman:1973jx}
\be\label{eq:0CWdef}
V_{\rm CW}=\frac{1}{64\pi^2}{\rm STr}{\cal M}^4\ln\nbrack{\frac{{\cal M}^2}{\Lambda^2}}=\frac{1}{64\pi^2}\sum\sbrack{m_B^4\ln\nbrack{\frac{m_B^2}{\Lambda^2}}-m_f^4\ln\nbrack{\frac{m_f^2}{\Lambda^2}}}
\ee
which evaluates to
\begin{align}
V_{\rm CW}= & \frac{nh^4\mu^4}{16\pi^2}\ln\nbrack{\frac{h^2|\hat{\Phi}|^2}{\Lambda^2}}+\nonumber\\
& \frac{nh^4|\hat{\Phi}|^4}{32\pi^2}\sbrack{\nbrack{1+\frac{\mu^2}{|\hat{\Phi}|^2}}^2\ln\nbrack{1+\frac{\mu^2}{|\hat{\Phi}|^2}}+\nbrack{1-\frac{\mu^2}{|\hat{\Phi}|^2}}^2\ln\nbrack{1-\frac{\mu^2}{|\hat{\Phi}|^2}}}
\end{align}
for each diagonal meson component $\hat{\Phi}$.  For $|\hat{\Phi}|\gg\mu$ it can be approximated by
\be\label{eq:0CWapprox}
V_{\rm CW}\approx\frac{nh^4\mu^4}{32\pi^2}\sbrack{3+2\ln\nbrack{\frac{h^2|\hat{\Phi}|^2}{\Lambda^2}}}\,.
\ee
This function has no minimum, ergo $|\hat{\Phi}|$ cannot be stabilised with $|\Phi|\gg\mu$.  Actually, this approximation becomes valid even when $|\hat{\Phi}|$ is only a little higher than $\mu$ so we find that $\Phi$ cannot be stabilised at a suitable value at all.  This problem was discussed in more detail in \cite{Giveon:2009yu}, where the solution was to add a second quark mass scale to the electric theory $m_Q\rightarrow m_1,m_2\implies\mu\rightarrow\mu_1,\mu_2$.  A stable vacuum can then be found when $|\hat{\Phi}|$ lies between the two scales.

\section{Baryon deformations in $\SU{2}$\label{sec:BD}}

If we want to stabilise the diagonal components of the pseudomodulus $\Phi$ ($\hat{\Phi}$) by adding an operator to the superpotential there are two simple options\footnote{Other simple deformations would have no equivalent in the electric theory so are less well motivated.}.  The first is to add a meson deformation to the superpotential of \eqref{eq:ISSWmg}:
\be
\frac{1}{h}W_{\rm mg}\longrightarrow\frac{1}{h}W_{\rm mg}+\frac{1}{h}f(\hat{\Phi})
\ee
where $f(\hat{\Phi})$ is a polynomial of order $r$.  Such a function can easily be generated in the electric theory.  This deformation would lead to several supersymmetric vacua.  Specifically, the $F$-term for $\hat{\Phi}$ becomes
\be
F_{\hat{\Phi}}=h\nbrack{f^{\prime}(\hat{\Phi})-\mu^2}
\ee
at $q=\tilde{q}^T=0$.  The expression on the right hand side is an order $r-1$ polynomial so has $r-1$ roots and thus leads to $r-1$ supersymmetric vacua.  If we desire a non-supersymmetric vacuum we must find a stationary point of the one-loop potential $V_{\rm 1-loop}(\hat{\Phi})$ which is \emph{not} a solution to $F_{\hat{\Phi}}=0$.  This is only possible if $V_{\rm 1-loop}(\hat{\Phi})$ has more than $r-1$ minima.  Using the expression \eqref{eq:0CWapprox} for the tachyon  free regime we find
\begin{align}
& V_{\rm 1-loop}(\hat{\Phi})\sim\ln\hat{\Phi}+\nbrack{f^{\prime}(\hat{\Phi})}^2\nonumber\\
\implies & V_{\rm 1-loop}^{\prime}(\hat{\Phi})\sim\frac{1}{\hat{\Phi}}+f^{\prime\prime}(\hat{\Phi})f^{\prime}(\hat{\Phi})\,.
\end{align}
$V_{\rm 1-loop}^{\prime}(\hat{\Phi})$ is therefore an order $2(r-1)$ polynomial which has $2(r-1)$ roots.  No more than half of these can be minima, leaving at most $r-1$ minima; exactly the same as the number of supersymmetric minima.  We thus conclude that a non-supersymmetric vacuum with $q=\tilde{q}^T=0$ cannot be stabilised in this way.

The second option is to deform the superpotential with a baryonic deformation\footnote{Baryon deformations can easily be generated in the electric theory by using the baryon map \eqref{eq:ISSBmap}}.  Although baryonic deformations were discussed in a similar context in \cite{Abel:2007jx} the vacua studied here are not the same.  We will be interested in vacua which are uplifted with respect to those in \cite{Abel:2007jx} and hence have very different phenomenological properties.  Baryonic deformations do not include $\Phi$ explicitly so can only stabilise it via the Coleman-Weinberg potential.  As such, we restrict ourselves to the magnetic gauge group $\SU{2}$ where baryonic deformation look like mass terms.  The superpotential \eqref{eq:ISSWmg} is deformed to
\be\label{eq:BDdef}
\frac{1}{h}W_{\rm mg}\longrightarrow\tilde{q}\Phi q-\mu^2\Phi+m_q\epsilon^{(2)}q_1q_2+\tilde{m}_q\epsilon_{(2)}\tilde{q}^1\tilde{q}^2
\ee
where $m_q$ and $\tilde{m}_q$ are dimension 1 coupling constants and $\epsilon^{(2)}$ represents contraction over the colour indices with a rank 2 alternating tensor.  The subscripts on the quarks (superscripts on the antiquarks) denote flavour indices so we see that this deformation explicitly breaks the global symmetry group, from $\SU{N_f}\times\U{1}_B$ down to $\SU{2}\times\SU{N_f-2}$.  The other flavour and colour indices have now been suppressed.

The baryonic deformation retains the classical stationary point at $q=\tilde{q}^T=0$ and does not change the status of $\Phi$ as a pseudomodulus.  However, breaking the flavour symmetry forces us to expand around this vacuum in components
\be\label{eq:BDexpan}
q=\nbrack{\begin{array}{c} x \\ y \end{array}},\quad
\tilde{q}=\nbrack{\begin{array}{cc} \tilde{x} & \tilde{y} \end{array}},\quad
\Phi=\nbrack{\begin{array}{cc} \phi & \tilde{\rho} \\ \rho & \chi \end{array}}
\ee
where $x$ and $\phi$ are $2\times2$ matrices, $y$ and $\rho$ are $(N_f-2)\times2$ matrices and $\chi$ is a $(N_f-2)\times(N_f-2)$ matrix (with transposed relations for letters with a tilde).  The baryon deformation only affects the $\SU{2}$ flavour sector (we will discuss the remainder of the theory later) and we can again use the residual flavour symmetry to restrict our attention to the diagonal components of $\Phi$.  The relevant superpotential terms are therefore
\be
\frac{1}{h}W_{\rm mg}\supset\tilde{x}\phi x-\mu^2\phi+m_q\epsilon^{(2)}x_1x_2+\tilde{m}_q\epsilon_{(2)}\tilde{x}^1\tilde{x}^2\,.
\ee
For this superpotential, the mass squared eigenvalues for combinations of the $x$'s are
\begin{align}\label{eq:BDm2}
\hat{m}_B^2= & \frac{h^2}{2}\nbrack{m_+^2+m_-^2+|\phi_+|^2+|\phi_-|^2}\pm\nonumber\\
& h^2\sqrt{\nbrack{m_+^2+|\phi_-|^2}\nbrack{m_-^2+|\phi_+|^2}+\mu^4\pm2\mu^2|\phi_+|\sqrt{m_+^2+|\phi_-|^2}}\nonumber\\
\hat{m}_f^2= & \frac{h^2}{2}\nbrack{m_+^2+m_-^2+|\phi_+|^2+|\phi_-|^2\pm2\sqrt{\nbrack{m_+^2+|\phi_-|^2}\nbrack{m_-^2+|\phi_+|^2}}}
\end{align}
where the $\pm$'s are independent and we have defined
\be
\phi_{\pm}=\frac{1}{\sqrt{2}}\nbrack{\phi{}^1_1\pm \phi{}^2_2}\,,\quad
m_{\pm}=\frac{1}{\sqrt{2}}\nbrack{m_q\pm\tilde{m}_q}\,.
\ee
The $\phi$'s are pseudomoduli so the stationary point $x=\tilde{x}=\phi=0$ has no tree level tachyons only if
\be\label{eq:BDnotach}
m_+^2+m_-^2>2\sqrt{m_+^2m_-^2+\mu^4}\,.
\ee

\begin{figure}[!th]
\begin{center}
\includegraphics[width=7.4cm]{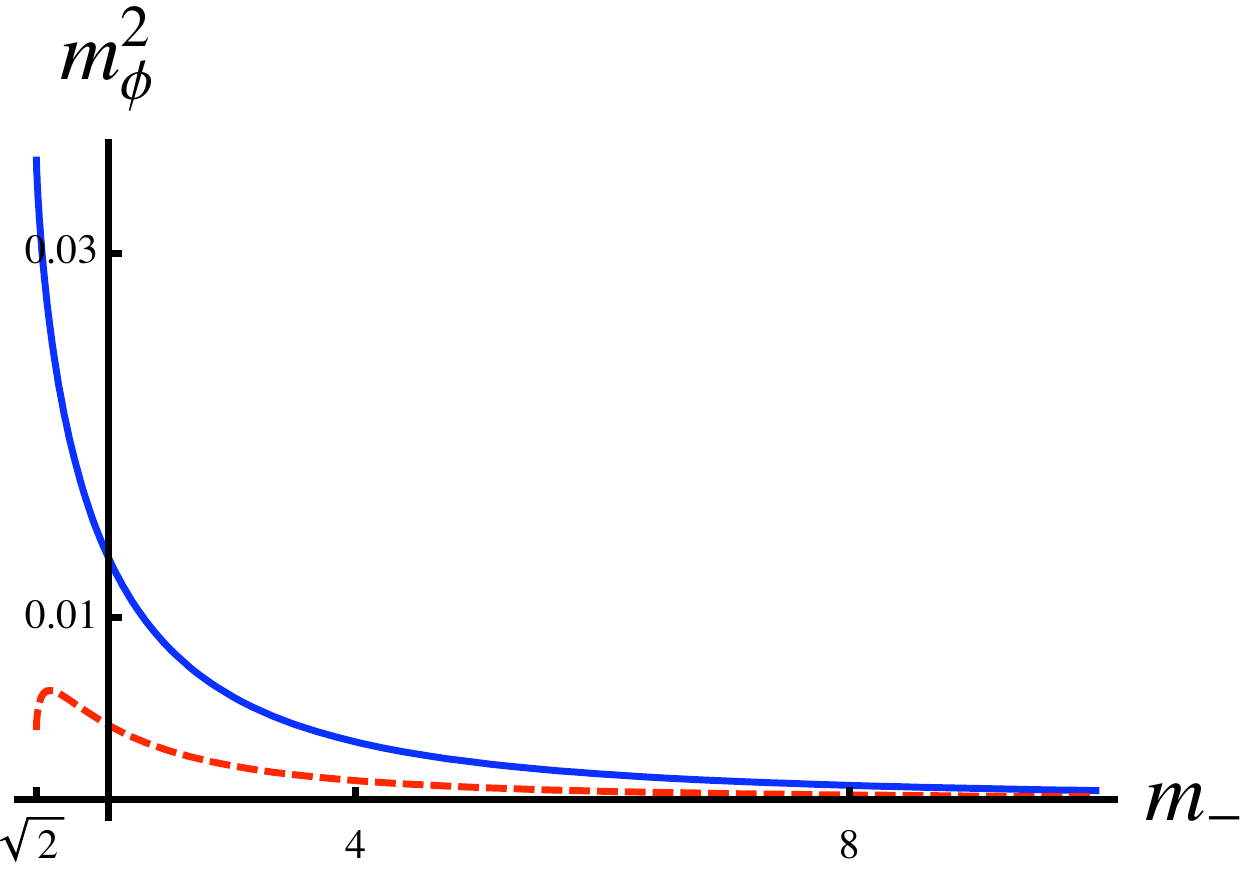}
\includegraphics[width=7.4cm]{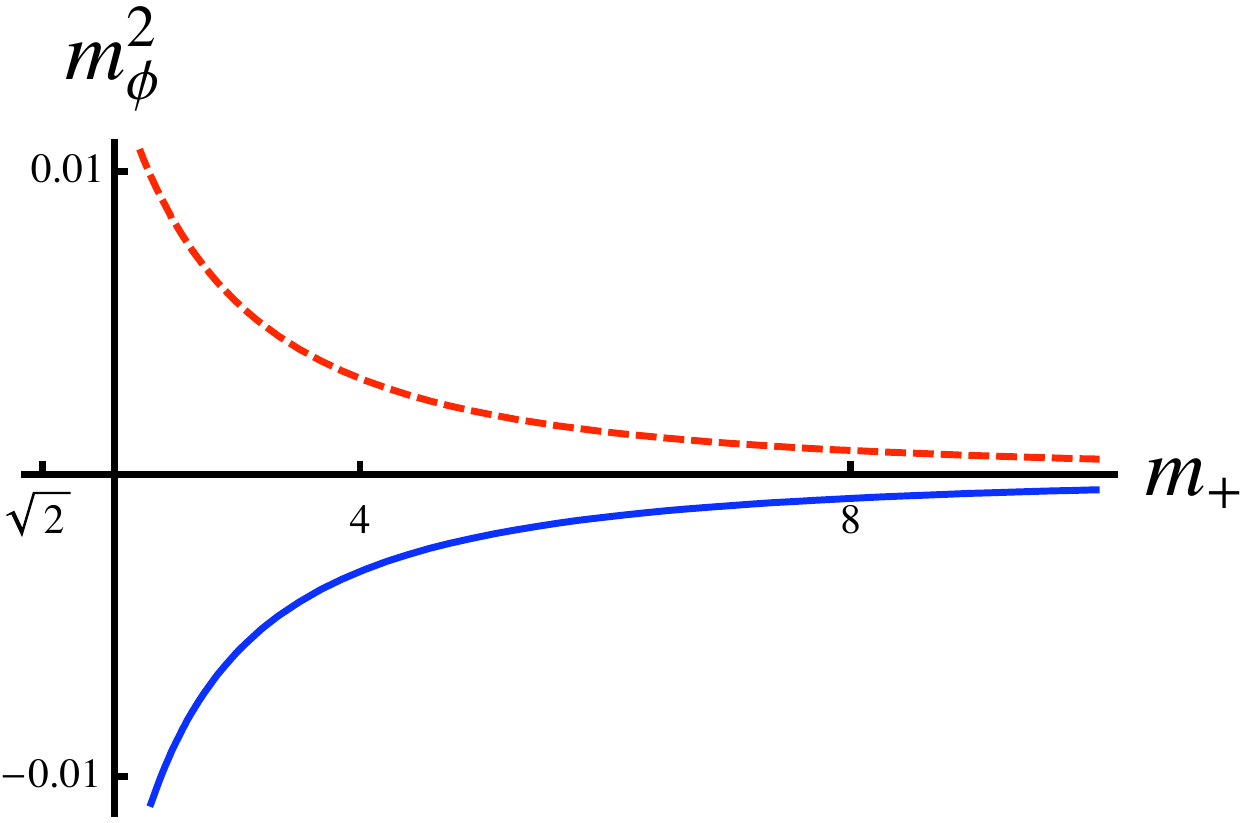}
\caption{\emph{The masses of the pseudomoduli $\phi_{\pm}$ as functions of $m_-$ when $m_+=0$ (left) and $m_+$ when $m_-=0$ (right).  The solid line shows the mass-squared of $\phi_+$ and the dashed line $\phi_-$.  $m_{\pm}$ are in units of $\mu$ and $m_{\phi}^2$ is in units of $h^4\mu^2$.}\label{fig:BDphimass}}
\end{center}
\end{figure}

The simplified cases $m_+=0$ and $m_-=0$ can be investigated analytically without too much trouble.  It is then possible to numerically interpolate between these extreme examples.  First, consider the case $\tilde{m}_q=-m_q$, i.e.\ $m_+=0$.  If we calculate the Coleman-Weinberg potential \eqref{eq:0CWdef} we can deduce the mass-squareds acquired by the pseudomoduli at one-loop.  They are
\begin{align}
m_{\phi_+}^2= & \frac{h^4\mu^2}{8\pi^2}\sbrack{\nbrack{\frac{m_-^2}{\mu^2}+2}\ln\nbrack{\frac{m_-^2}{\mu^2}+2}+\nbrack{\frac{m_-^2}{\mu^2}-2}\ln\nbrack{\frac{m_-^2}{\mu^2}-2}-\frac{m_-^2}{\mu^2}\ln\nbrack{\frac{m_-^4}{\mu^4}}}\nonumber\\
m_{\phi_-}^2= & \frac{h^4\mu^2}{8\pi^2}\left[\nbrack{3\frac{m_-^2}{\mu^2}+\frac{m_-^4}{\mu^4}+2}\ln\nbrack{\frac{m_-^2}{\mu^2}+2}+\nbrack{3\frac{m_-^2}{\mu^2}-\frac{m_-^4}{\mu^4}-2}\ln\nbrack{\frac{m_-^2}{\mu^2}-2}-\right.\nonumber\\
& \left.4\frac{m_-^2}{\mu^2}\nbrack{1-\ln\nbrack{\frac{m_-^3}{\mu^3}}}\right]\,.
\end{align}
One immediately sees that we require $m_-^2>2\mu^2$ for these masses to be well defined.  This is equivalent to \eqref{eq:BDnotach} evaluated at $m_+=0$.  The behaviour of these masses is shown in Figure \ref{fig:BDphimass}: we find that both pseudomoduli acquire positive mass-squareds and are stabilised at the origin.  Consequently a stable vacuum exists at the origin with the mass-squared eigenvalues in the $x$-sector evaluating to
\be
\hat{m}_B^2=h^2\nbrack{\frac{1}{2}m_-^2\pm\mu^2}\,,\quad\hat{m}_f^2=\frac{1}{2}h^2m_-^2\,.
\ee

Meanwhile for $\tilde{m}_q=m_q$, i.e.\ $m_-=0$ the mass-squareds for the pseudomoduli are
\begin{align}
m_{\phi_+}^2= & \frac{h^4\mu^2}{8\pi^2}\sbrack{\nbrack{3\frac{m_+^2}{\mu^2}+2}\ln\nbrack{\frac{m_+^2}{\mu^2}+2}+\nbrack{3\frac{m_+^2}{\mu^2}-2}\ln\nbrack{\frac{m_+^2}{\mu^2}-2}-4\frac{m_+^2}{\mu^2}\ln\nbrack{\frac{m_+^3}{\mu^3}}}\nonumber\\
m_{\phi_-}^2= & \frac{h^4\mu^2}{8\pi^2}\sbrack{\nbrack{\frac{m_+^2}{\mu^2}+2}\ln\nbrack{\frac{m_+^2}{\mu^2}+2}+\nbrack{\frac{m_+^2}{\mu^2}-2}\ln\nbrack{\frac{m_+^2}{\mu^2}-2}-\frac{m_+^2}{\mu^2}\ln\nbrack{\frac{m_+^4}{\mu^4}}}\,.
\end{align}
We require $m_+^2>2\mu^2$ (again, \eqref{eq:BDnotach} evaluated at $m_-=0$) for these masses to be well defined.  However, $\phi_+$ now acquires a negative mass-squared so the origin is unstable.  This is also shown in Figure \ref{fig:BDphimass}.

\begin{figure}[!th]
\begin{center}
\includegraphics[width=7.4cm]{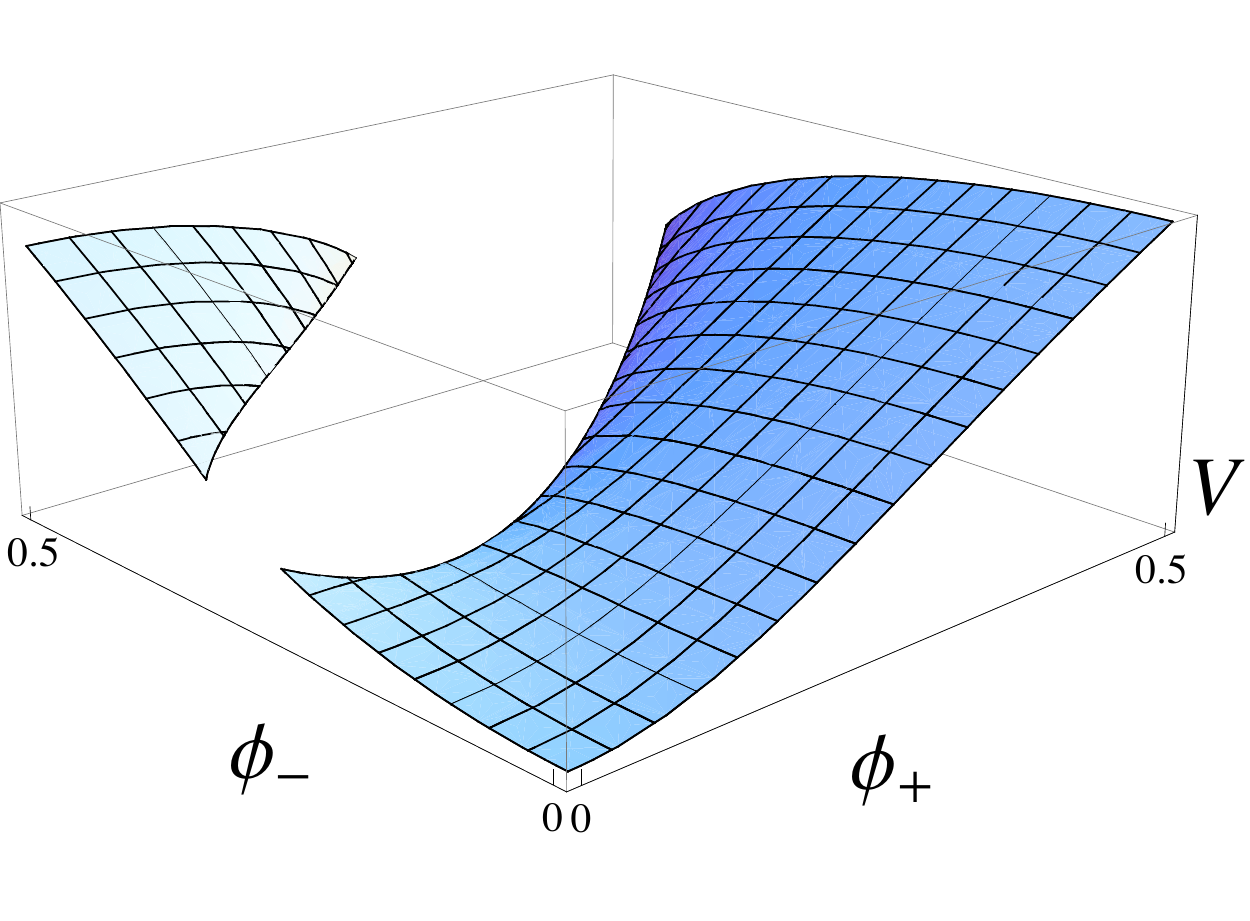}
\includegraphics[width=7.4cm]{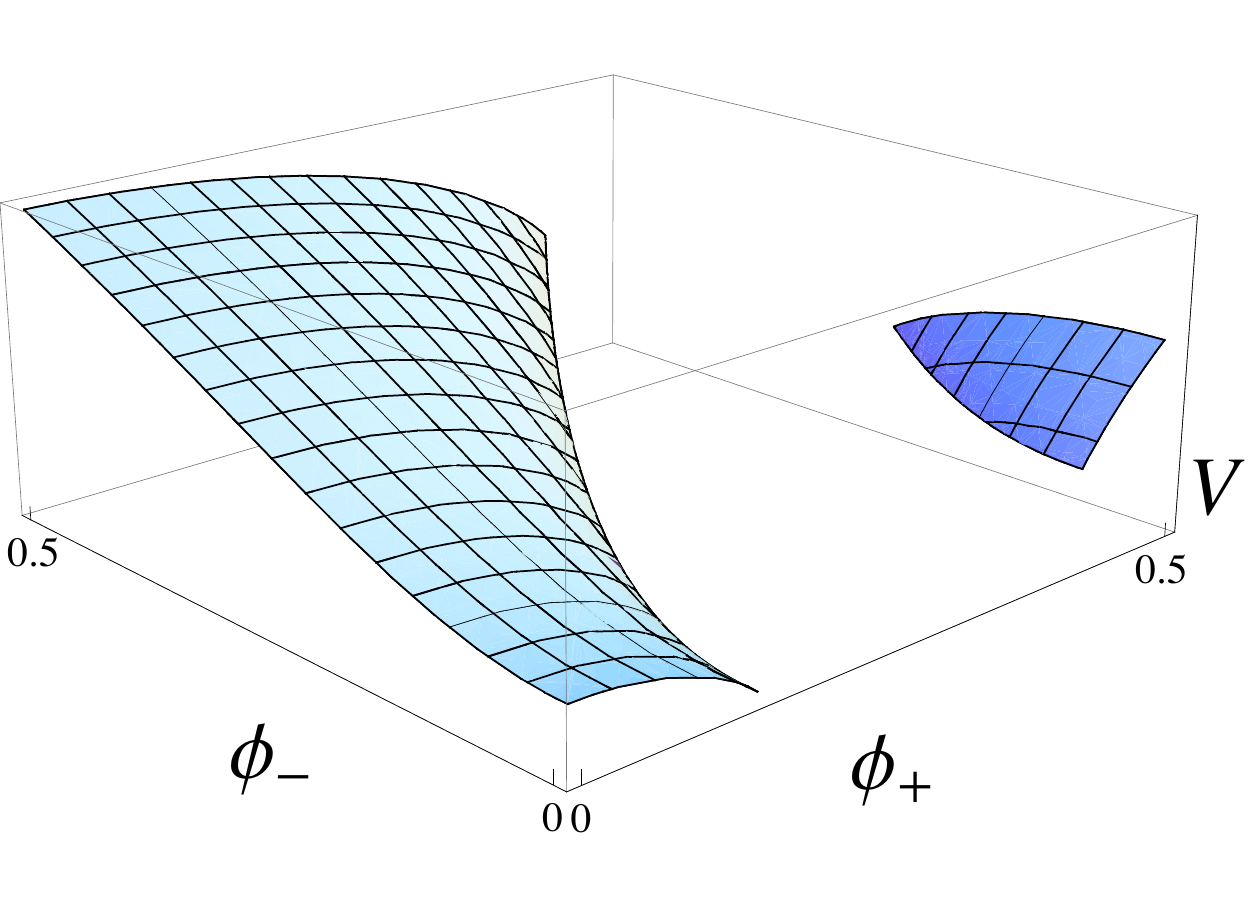}
\caption{\emph{The one-loop potential as a function of the pseudomoduli $\phi_{\pm}$ (assumed real) when $m_+=0$ and $m_-\neq0$ (left) and $m_-=0$ and $m_+\neq0$ (right).  It is symmetric about the origin in $\phi_{\pm}$ for both cases.  Gaps in the surface show regions containing tree level tachyons among the $x$ fields, where the theory presumably runs to the vacuum similar to that of \cite{Abel:2007jx}.  Parameters have been set to $\Lambda=1$, $h=1$, $\mu=1/10$ and $m_{\pm}=\sqrt{2}/5$ when not equal to zero.}\label{fig:BDV1l}}
\end{center}
\end{figure}

A surface plot of the full one-loop potential as a function of the pseudomoduli is shown in Figure \ref{fig:BDV1l}.  Both cases contain tree level tachyons somewhere in the pseudomoduli space (where the one-loop potential cannot be defined) but when $m_+=0$ and $m_-\neq0$ we see both $\phi_+$ and $\phi_-$ are stabilised around zero.  The $\SU{2}$ flavour symmetry remains unbroken.  To get a handle on where the tachyons appear we can set $\phi_+=m_+=0$ in \eqref{eq:BDm2} to find
\be
\hat{m}_B^2=\frac{h^2}{2}\nbrack{m_-^2+|\phi_-|^2\pm2\sqrt{m_-^2|\phi_-|^2+\mu^4}}
\ee
which becomes negative when
\be
m_-^2-2\mu^2<|\phi_-|^2<m_-^2+2\mu^2\,.
\ee
Increasing $m_-$ therefore moves the tachyonic regions further away from $|\phi_-|=0$, while increasing $\mu$ makes them wider.  In the vacuum at the origin the Coleman-Weinberg potential evaluates to
\be
V_{\rm CW}=\frac{h^4}{32\pi^2}\sbrack{4\mu^4\ln\nbrack{\frac{h^4}{4\Lambda^4}\sbrack{m_-^4-4\mu^4}}+m_-^4\ln\nbrack{1-4\frac{\mu^4}{m_-^4}}+4m_-^2\mu^2\ln\nbrack{\frac{m_-^2+2\mu^2}{m_-^2-2\mu^2}}}\,.
\ee
On the other hand, when $m_-=0$ and $m_+\neq0$, $\phi_+$ runs away from the origin to a region containing tachyons among the $x$ fields so this stationary point cannot be stable.  In either case, we expect the theory to run to a vacuum similar to the one discussed in \cite{Abel:2007jx} once it strays into a tachyonic region.

\section{Gauging the flavour group\label{sec:G}}

If we are to use this model as a SUSY breaking sector in a direct mediation scenario, gauging the flavour group is a natural thing to do.  In this case, $\phi_-$ is no longer a pseudomodulus when $x=\tilde{x}=0$ as it is not a $D$-flat direction; $D$-terms automatically stabilise it at $\phi_-=0$.  This opens up a new possibility for the deformed superpotential \eqref{eq:BDdef} that can yield a stable, uplifted vacuum: setting $\tilde{m}_q=0$ (or equivalently we could set $m_q=0$).

\begin{figure}[!th]
\begin{center}
\includegraphics[width=7.4cm]{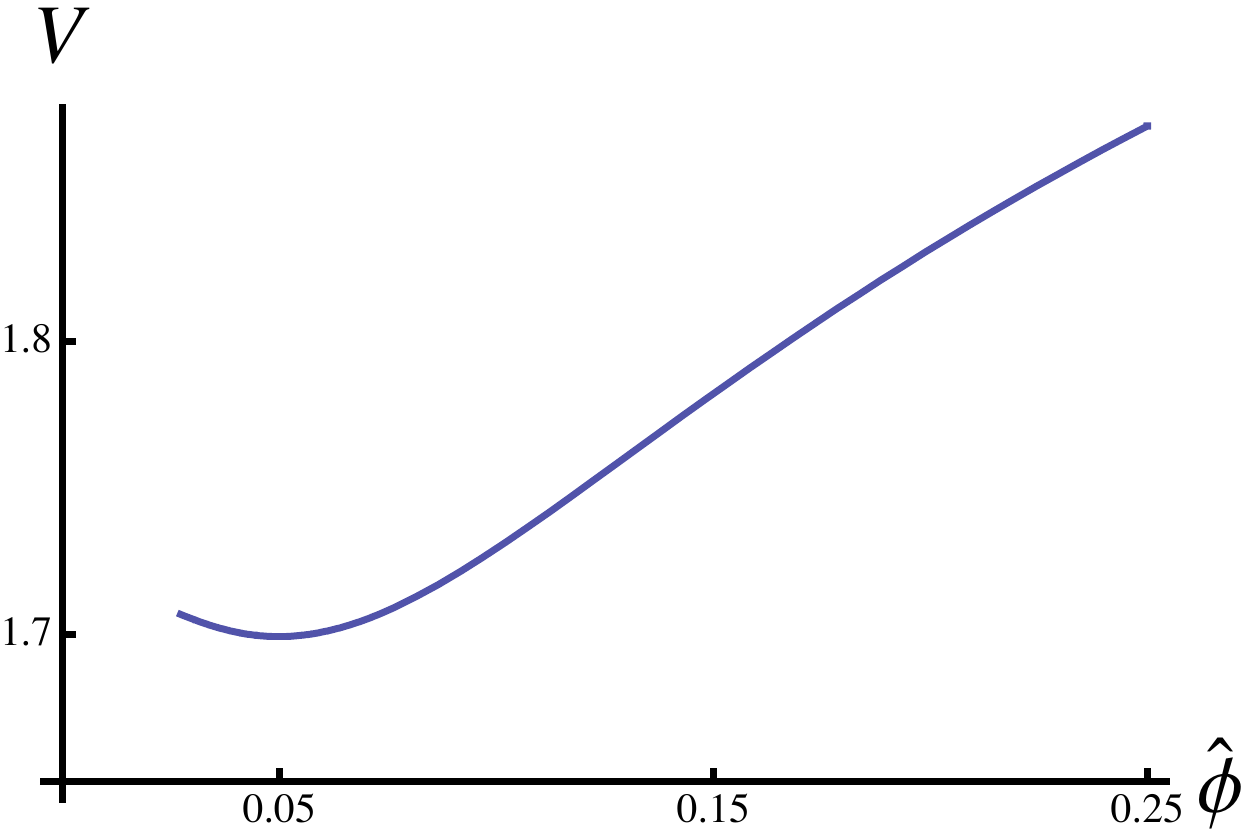}
\caption{\emph{The one-loop potential as a function of the pseudomodulus $\hat{\phi}$ (assumed real) when $\tilde{m}_q=0$ and the flavour symmetry is gauged.  Parameters have been set to $\Lambda=1$, $h=1$, $\mu=1/200$ and $m_q=1/5$.  The potential is given in units of $\mu^4$.  Note the minimum at $\hat{\phi}\approx m_q/4$.  Increasing $m_q$ moves the minimum to the right, whereas increasing $\mu$ moves the region containing tree level tachyons (where the potential is not defined) further right.}\label{fig:GV1l}}
\end{center}
\end{figure}

We can go through the process of finding the mass-squard eigenstates and calculating the Coleman-Weinberg potential as before.  This is now a function of the one remaining pseudomodulus, $\phi_+$, and can be easily deduced from \eqref{eq:BDm2} by setting $m_+=m_-=m_q/\sqrt{2}$, $\phi_-=0$ and redefining $\phi_+=\sqrt{2}\hat{\phi}$ (where $\phi=\hat{\phi}\uno_2$).  For $\mu^2\ll m_q^2$ we find
\begin{align}\label{eq:GVCW}
V_{\rm CW}\approx\frac{h^4\mu^4}{2\pi^2\nbrack{m_q^2+4|\hat{\phi}|^2}} & \left[m_q^2+12|\hat{\phi}|^2+\nbrack{m_q^2+4|\hat{\phi}|^2}\ln\nbrack{\frac{h^4|\hat{\phi}|^4}{\Lambda^4}}\right.+\nonumber\\
& \left.m_q\frac{m_q^2+2|\hat{\phi}|^2}{\sqrt{m_q^2+4|\hat{\phi}|^2}}\ln\nbrack{\frac{m_q^2+2|\hat{\phi}|^2+m_q\sqrt{m_q^2+4|\hat{\phi}|^2}}{m_q^2+2|\hat{\phi}|^2-m_q\sqrt{m_q^2+4|\hat{\phi}|^2}}}\right]\,.
\end{align}
This function can be reduced to a function of $|\hat{\phi}|/m_q$ then shown numerically to have a minimum at $|\hat{\phi}|\approx m_q/4$ for any values of $h$, $\Lambda$ and $\mu^2\ll m_q^2$.  This minimum provides a new uplifted vacuum for the theory.  In it we find
\be\label{eq:GVCWmin}
V_{\rm CW}\approx\frac{h^4\mu^4}{\pi^2}\sbrack{3+2\ln\nbrack{\frac{hm_q}{4\Lambda}}}
\ee
and the mass-squared of $\hat{\phi}$ goes like $0.63h^4\mu^4/m_q^2$.  We find from \eqref{eq:BDm2} that there are no tachyons in the $x$-sector as long as $|\hat{\phi}|^3\gtrsim4m_q\mu^2$.  For the minimum at $|\hat{\phi}|\approx m_q/4$ to fall in this regime we require $m_q\gtrsim16\mu$.  A typical example of the potential is plotted in Figure \ref{fig:GV1l}.  Despite the fact that $\phi$ acquires a VEV in this vacuum, it is only the trace that is non-zero.  Consequently, the $\SU{2}$ symmetry remains unbroken.

\section{Direct mediation\label{sec:DM}}

In the model discussed above there always exist tachyonic directions at tree level somewhere in the pseudomoduli space of the theory.  According to \cite{Komargodski:2009jf} this may allow us to get around the problem of anomalously small gaugino masses common to theories of direct mediation.  We will now discuss three possible approaches for using this model in a direct mediation scenario.  We choose $m_q\neq0$ and $\tilde{m}_q=0$ in both sections so the results from \S\ref{sec:G} apply.

\subsection{$\SU{N_f-2}$ mediation\label{sec:Nf-2}}

The most obvious choice is to gauge the entire flavour group and embed the visible sector gauge group in the $\SU{N_f-2}$ part of the flavour symmetry.  Unfortunately this sector is decoupled from the SUSY breaking in the $\SU{2}$ flavour sector.  To rectify this we must include a coupling from $\phi$ to either the $y$'s or the $\rho$'s of \eqref{eq:BDexpan}.  Generating couplings to the quarks in the magnetic theory using operators in the electric theory is difficult so we will use the $\rho$'s instead.  An appropriate coupling can then be generated by a cubic meson operator, giving
\be\label{eq:DMW}
\frac{1}{h}W_{\rm mg}=\tilde{q}\Phi q-\mu^2\Phi+m_q\epsilon^{(2)}x_1x_2+\eta\Tr{}{\rho\phi\tilde{\rho}}+\frac{1}{2}m_{\chi}\Tr{}{\chi^2}\,.
\ee
We have also added a mass term for $\chi$ which stabilises it at $\mu^2/m_{\chi}$; without this term, $\chi$ would be unstable for the same reasons as $\hat{\Phi}$ in \S\ref{sec:UpV}.  Both deformations correspond to non-renormalisable operators in the electric theory, but we will motivate their presence in \S\ref{sec:PV} using arguments similar to \cite{Murayama:2006yf}.  In brief, the magnetic deformations considered here will correspond to \emph{all} generic deformations up to dimension six in the electric theory that are compatible with the symmetries.  Higher dimension operators will be parametrically suppressed in the magnetic theory so can be safely ignored.  Even so, we could in principle add other operators arising from $\Phi^2$ or $\Phi^3$ deformations without breaking the residual flavour symmetry.  Those not involving $\phi$ do not change our results significantly\footnote{In fact, they only improve the situation by pushing the masses of the $\rho$'s higher so that Landau poles pose even less of a problem.}, but those that do could have an adverse effect.

Specifically we do not want the operators $\phi^2$, $\phi\chi$, $\phi^3$, $\phi^2\chi$ and $\phi\chi^2$ to appear.  The mixed $\phi$-$\chi$ operators correspond to multitrace deformations in the electric theory.  If we restrict our attention to single trace operators they can be discarded.  This can be motivated by assuming the electric theory is itself a low energy effective theory (as it must be to explain the presence of non-renormalisable operators) embedded in a theory of intersecting NS and D-branes.  Such theories were studied in \cite{Giveon:2007ew} where it was shown that single trace operators can be generated naturally whereas multitrace operators are not.  To see why the remaining terms are not allowed, imagine restoring SUSY by setting $\mu=0$.  An anomalous $R$-symmetry is also restored with
\begin{align}
R(x) & =1 & R(y) & =R_{\rho} & R(\rho) & =R_{\rho} & R(\phi) & =\frac{1}{2} \nonumber\\
R(\tilde{x}) & =\frac{1}{2} & R(\tilde{y}) & =1-R_{\rho} & R(\tilde{\rho}) & =\frac{3}{2}-R_{\rho} & R(\chi) & =1
\end{align}
forbidding both $\phi^2$ and $\phi^3$ at tree level.  This $R$-symmetry explicitly breaks the flavour symmetry of the undeformed model, meaning we can't consider the components $\phi$, $\chi$, $\rho$ and $\tilde{\rho}$ as part of a single field $\Phi$.  However, the baryon deformation already breaks the symmetry in this way -- $\Phi$ is used in \eqref{eq:DMW} merely for notational brevity.  We expect the $R$-symmetry to be broken in a non-supersymmetric, metastable vacuum anyway due to the arguments of \cite{Nelson:1993nf} but now both the $R$-symmetry breaking and the SUSY breaking are described by a single parameter, $\mu$.

\begin{figure}[!th]
\begin{center}
\includegraphics[width=7.4cm]{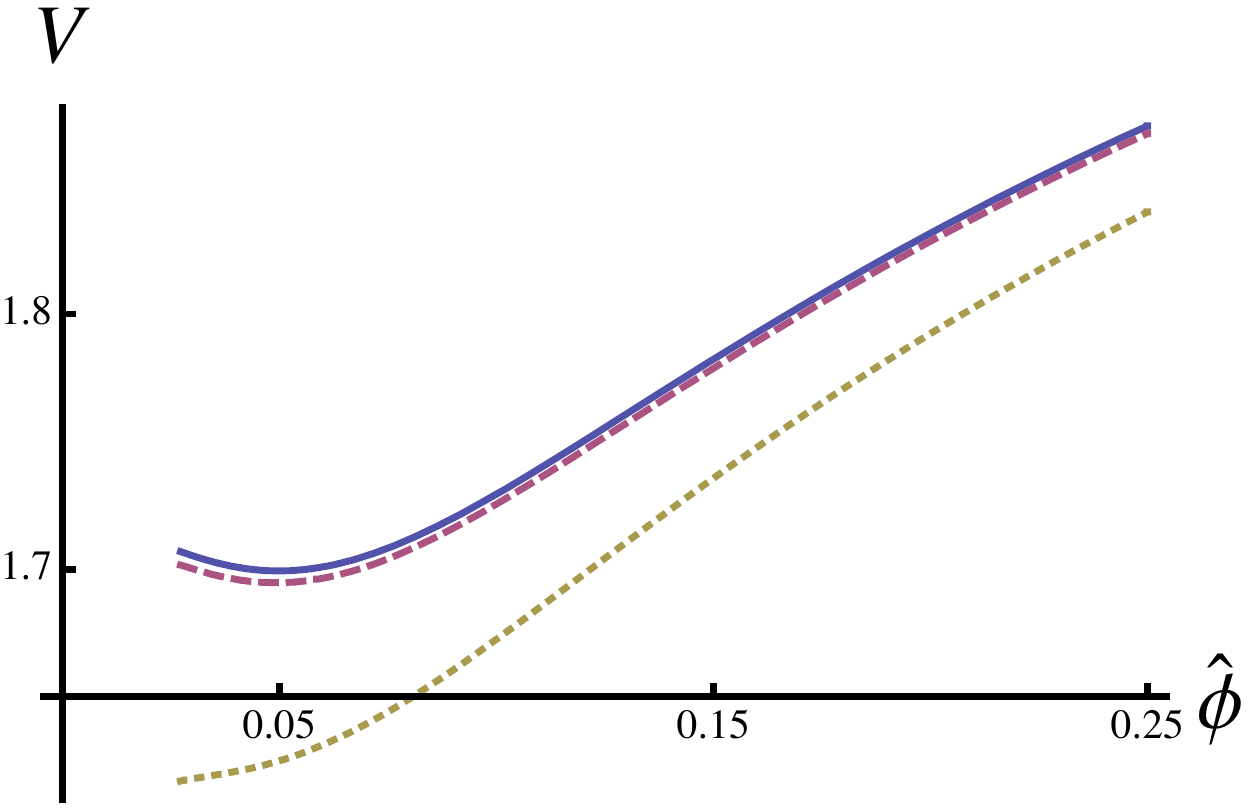}
\caption{\emph{The one-loop potential as a function of the pseudomodulus $\hat{\phi}$ (assumed real) in the $\SU{N_f-2}$ direct mediation scenario.  Parameters have been set to $\Lambda=1$, $h=1$, $\mu=1/200$, $m_q=1/5$.  The potential is given in units of $\mu^4$.  The solid line shows $\eta=0$, the dashed line $\eta=1/10$ and the dotted line $\eta=1/2$.  The minimum at $\hat{\phi}\approx m_q/4$ is robust to the inclusion of $\eta$, for $\eta\lesssim1/3$.}\label{fig:Nf-2V1l}}
\end{center}
\end{figure}

The tree level vacuum resulting from \eqref{eq:DMW} is
\be
\vac{N_f-2}:\quad\quad q=\nbrack{\begin{array}{c} 0 \\ 0 \end{array}},\quad
\tilde{q}=\nbrack{\begin{array}{cc} 0 & 0 \end{array}},\quad
\Phi=\nbrack{\begin{array}{cc} \hat{\phi}\uno_2 & 0 \\ 0 & \mu^2/m_{\chi} \end{array}},\quad
V_{\rm tree}=2h^2\mu^4\,.
\ee
with the fields expanded as in \eqref{eq:BDexpan}.  The $\rho$'s now act as messengers, coupling the SUSY breaking in the $\SU{2}$ flavour sector to the $\SU{N_f-2}$ flavour sector containing the visible sector gauge group.  The tree level mass-squared eigenvalues in the $x$-sector are unchanged from \S\ref{sec:G}.  Fluctuations of the remaining scalar fields acquire tree level mass-squareds
\begin{align}\label{eq:DMmass2}
y,\tilde{y}\quad: & \quad \hat{m}_B^2=h^2\mu^4/m_{\chi}^2\nonumber\\
\chi\quad: & \quad \hat{m}_B^2=h^2m_{\chi}^2\nonumber\\
\rho\pm\tilde{\rho}^{\dag}\quad: & \quad \hat{m}_B^2=h^2\nbrack{\eta^2|\hat{\phi}|^2\mp2\eta\mu^2}\,.
\end{align}
We thus require $|\hat{\phi}|^2>2\mu^2/\eta$ for the $\rho$'s to be non-tachyonic and so achieve a stable vacuum.  The effect on the Coleman-Weinberg potential is to add an extra term to \eqref{eq:GVCW}
\be
\Delta V_{\rm CW}=\frac{h^4\eta^2\mu^4}{4\pi^2}\sbrack{3+4\ln\nbrack{\frac{h^4\eta^4|\hat{\phi}|^4}{\Lambda^4}}}\,.
\ee
As long as $\eta^2\lesssim1/3$, $\Delta V_{\rm CW}\lesssim10^{-1}V_{\rm CW}$ so the perturbation does not destroy the minimum in the potential at $|\hat{\phi}|\approx m_q/4$.  Combined with the tachyon constraint $|\hat{\phi}|^2>2\mu^2/\eta$ evaluated at $|\hat{\phi}|=m_q/4$ minimum we find
\be
\frac{32\mu^2}{m_q^2}\lesssim\eta\lesssim\frac{1}{3}\,.
\ee
As $m_q\gtrsim16\mu$ the left hand side can be at most $1/8$ so this range of $\eta$ is quite reasonable.  Plots of the one-loop potential for various values of $\eta$ are given in Figure \ref{fig:Nf-2V1l}.

We can now calculate the gaugino mass via \cite{Cheung:2007es}
\be
m_{\lambda}\sim\Lambda_G=F_{\hat{\phi}}\pdiff{}{\hat{\phi}}\ln\det\nbrack{m_{\rm mess}}=\frac{2h\mu^2}{|\hat{\phi}|}
\ee
and the sfermion masses\footnote{More precisely $m_{\lambda}=C_{\lambda}\Lambda_G$ and $m_{\tilde{f}}^2=C_{\tilde{f}}^2\Lambda_S^2$, but $C_{\lambda}\sim C_{\tilde{f}}$ so these coefficients are unimportant when comparing the masses.} from
\be
m_{\tilde{f}}^2\sim\Lambda_S^2=\frac{1}{2}F_{\hat{\phi}}^2\frac{\partial^2}{\partial\hat{\phi}\hat{\phi}^*}\sum\sbrack{\ln\nbrack{\hat{m}_{\rm mess}^2}}^2=\frac{32h^2\mu^4}{|\hat{\phi}|^2}
\ee
where $m_{\rm mess}=\eta\hat{\phi}\uno_2$ is the messenger mass matrix with eigenvalues $\hat{m}_{\rm mess}=\eta\hat{\phi}$.  The ratio of the gaugino masses to sfermion masses is conveniently parameterised by defining the effective number of messengers
\be
\cN_{\rm eff}=\frac{\Lambda_G^2}{\Lambda_S^2}=\frac{1}{8}
\ee
which is less than one but still sufficiently large.  Around the vacuum at $|\hat{\phi}|\approx m_q/4$ for visible sector gauge coupling $\alpha_{\rm vis}$ we have
\be\label{eq:DMmasses}
m_{\lambda}\sim\frac{2\alpha_{\rm vis}h\mu^2}{\pi m_q}\,,\quad m_{\tilde{f}}\sim\frac{4\sqrt{2}\alpha_{\rm vis}h\mu^2}{\pi m_q}\,.
\ee

\subsection{$\SU{2}\times\SU{3}$ mediation\label{sec:23}}

A second possibility is to choose $N_f=5$.  The SUSY breaking sector then has an $\SU{2}\times\SU{3}$ flavour symmetry that can neatly be identified with the visible sector gauge group.  We can get away with the simpler superpotential
\be
\frac{1}{h}W_{\rm mg}=\tilde{q}\Phi q-\mu^2\Phi+m_q\epsilon^{(2)}x_1x_2+\frac{1}{2}m_{\chi}\Tr{}{\chi^2}\,.
\ee
Supersymmetry breaking occurs in the $\SU{2}$ sector and is now mediated to the visible sector by the $x$'s.  Mediation to the $\SU{3}$ sector occurs via visible sector interactions so is suppressed by an extra loop.  This approach leads to an interesting signature: $\SU{3}$ masses would be less than $\SU{2}$ masses by a factor of one loop.  However, this mass hierarchy would likely upset gauge coupling unification in the visible sector.

\subsection{$\Sp{N_f}$ mediation}

It is worth briefly mentioning a third possibility for direct mediation available in this model.  Rather than using a baryon deformation of the form
\be
\frac{1}{h}W_{\rm mg}\supset m_q\epsilon^{(2)}q_1q_2
\ee
that singles out two flavours of quark, we could generalise and use a deformation
\be
\frac{1}{h}W_{\rm mg}\supset m_q^{ij}\epsilon^{(2)}q_iq_j
\ee
where $m_q$ is now an antisymmetric matrix in flavour space.  This deformation explicitly breaks the flavour group from $\SU{N_f}$ to $\Sp{N_f}$, which we could then gauge.  We no longer have to expand the quarks as in \eqref{eq:BDexpan}; instead, the whole model behaves as the $\SU{2}$ flavour sector in \S\ref{sec:BD}, only with a rescaled potential $V\rightarrow\frac{N_f}{2}V$.  The only pseudomodulus behaves like $\hat{\phi}$ so is safely stabilised around $m_q/4$.  Unfortunately, it is difficult to embed the visible sector gauge group into $\Sp{N_f}$ so using this method in a direct mediation scenario is of limited practical use.

\section{Phenomenological viability\label{sec:PV}}

We now discuss some phenomenological aspects of the $\SU{N_f-2}$ direct mediation scenario of \S\ref{sec:Nf-2}.  For simplicity we will analyse the minimal case: take $N_f=7$ and embed the visible sector in an $\SU{5}$ GUT.  Regarding parameters in the superpotential \eqref{eq:DMW}, we have five: $h$, $\mu$, $m_q$, $\eta$ and $m_{\chi}$.  All have dimension 1 except $h$ and $\eta$, which are dimensionless.  $h$ is simply an order 1 coupling constant that translates directly from the electric theory.  The magnitudes of the other parameters can be estimated by looking at where they come from in the UV \cite{Murayama:2006yf}.  Using the baryon map \eqref{eq:ISSBmap} we have schematically
\ba
\mu^2\Phi & \leftrightarrow & m_Q\Tr{}{\tilde{Q}Q}\nonumber\\
m_q\epsilon^{(2)}q_1q_2 & \leftrightarrow & \frac{1}{M^2}\epsilon^{(5)}Q^5\nonumber\\
\eta\phi\tilde{\rho}\rho & \subset & \frac{1}{M^3}\Tr{}{(\tilde{Q}Q)^3}\nonumber\\
m_{\chi}\chi^2 & \subset & \frac{1}{M}\Tr{}{(\tilde{Q}Q)^2}
\ea
where $M$ is some high scale.  These operators are the only single trace, perturbative operators in the electric theory which are consistent with its symmetries (including the $R$-symmetry discussed in \S\ref{sec:Nf-2}) and of dimension six or below.  Higher dimension operators could exist but will be suppressed so can be discarded.  From dimensional arguments we expect
\be
\mu\sim\sqrt{m_Q\Lambda}\,,\quad
m_q\sim\Lambda\nbrack{\frac{\Lambda}{M}}^2\,,\quad
\eta\sim\nbrack{\frac{\Lambda}{M}}^3\,,\quad
m_{\chi}\sim\Lambda\nbrack{\frac{\Lambda}{M}}
\ee
and then, from \eqref{eq:DMmass2}\footnote{The fermionic components of these superfields acquire the same masses, except for the $\rho$'s whose fermions get masses $\sim\Lambda\nbrack{\frac{\Lambda}{M}}^4$},
\be
m_y\sim m_Q\nbrack{\frac{M}{\Lambda}}\,,\quad
m_{\rho}\sim\sqrt{\Lambda^2\nbrack{\frac{\Lambda}{M}}^8\pm m_Q\Lambda\nbrack{\frac{\Lambda}{M}}^3}\,.
\ee
In \S\ref{sec:G} and \S\ref{sec:Nf-2} we required
\be
m_q\gtrsim16\mu\sep{and}\frac{32\mu^2}{m_q^2}\lesssim\eta\lesssim\frac{1}{3}
\ee
for the vacuum to be stable, which corresponds to
\be\label{eq:PVineq1}
m_Q\lesssim10^{-9}\Lambda\sep{and}
\frac{\Lambda}{M}\lesssim10^{-1}\,.
\ee
The gaugino mass scale was calculated in \eqref{eq:DMmasses}.  For this to be around the weak scale (assuming $h\sim1$), we require
\be\label{eq:PVineq2}
m_{\lambda}\sim\frac{2\alpha_{\rm vis}h\mu^2}{\pi m_q}\sim10^{-2}m_Q\nbrack{\frac{M}{\Lambda}}^2\sim100\mbox{ GeV}\quad\implies\quad m_Q\sim\nbrack{\frac{\Lambda}{M}}^210^4\mbox{ GeV}\,.
\ee
Finally, for the theory to remain calculable we need mass contributions from the K\"ahler potential, which go like $F_{\hat{\phi}}^2/\Lambda^2=4h^2\mu^4/\Lambda^2$ \cite{Intriligator:2008fe}, to be smaller than masses generated at one loop, which go like $h^4\mu^4/m_q^2$.  This last constraint tells us that
\be
\frac{\Lambda}{M}\lesssim\sqrt{h}
\ee
so is already satisfied by \eqref{eq:PVineq1} when $h\sim1$.  By approximating the minimum in Figure \ref{fig:Nf-2V1l} with a triangular potential barrier we can make a rough estimate of the bounce action \cite{Duncan:1992ai}, and therefore the lifetime of the uplifted minimum.  We find
\be
S\sim\frac{\Delta\hat{\phi}^4}{V_{\rm min}}\sim\frac{\Lambda^2}{10^{10}\mbox{ GeV}^2}\nbrack{\frac{\Lambda}{M}}^4\nbrack{7+\frac{1}{\pi^2}\sbrack{3+4\ln\nbrack{\frac{\Lambda}{4M}}}}^{-1}
\ee
where $V_{\rm min}$ denotes the total value of the potential in the minimum, given by $7h^2\mu^4+V_{\rm CW}$ with $V_{\rm CW}$ evaluated as in \eqref{eq:GVCWmin}.  $\Delta\hat{\phi}\sim m_q/4$ is the tunnelling distance to the tachyonic region and therefore the width of the barrier.  The final step is accomplished using the expressions for $m_q$, $\mu$ and $m_Q$ derived in this section.  For $\Lambda/M\gg10^{-7}$ the bounce action is well approximated by $S\sim\Lambda^6/M^410^{11}\mbox{ GeV}^2$.

All of these constraints are consistent with one another and leave a wide range of choices for the three fundamental parameters input into the electric theory; $m_Q$, $\Lambda$ and $M$ (assuming all dimensionless couplings are of order 1).  If we minimise the number of scales by saturating \eqref{eq:PVineq1} and taking $\Lambda\sim10^{-1}M$, \eqref{eq:PVineq1} tells us that we will always require $m_Q\sim100\mbox{ GeV}$ and will achieve a stable vacuum for any high scale $M\gtrsim10^{12}\mbox{ GeV}$.  This is quite an attractive scenario as we effectively only require two scales: a high scale $M$ (such as the Planck scale or the GUT scale) and the weak scale of $100\mbox{ GeV}$.  It also results in a very large bounce action and consequently a long lived vacuum.  For the messengers $\rho$ and additional matter charged under the visible sector ($y$ and $\chi$) we find\footnote{A brief note on the decays of these particles: $\chi$ decays rapidly to $\tilde{y}y$ through couplings in the superpotential \eqref{eq:DMW}.  At first glance the fields $y$ and $\rho$ can only decay through mutual annihilation so we expect them to be fairly stable.  The mass of the $y$'s is less than 10 TeV which is insufficient for their relic density to overclose the universe \cite{Dimopoulos:1996gy}, but the mass of the $\rho$'s is large enough to cause cosmological problems.  Gravitational interactions can, however, induce extra superpotential terms that violate messenger number symmetry and allow decays to $y$'s or visible sector particles \cite{Jedamzik:2005ir}.}
\be\label{eq:PVmessmass}
m_y\sim10m_Q\sim10^3\mbox{ GeV}\,,\quad
m_{\rho}\sim10^{-5}M\gtrsim10^7\mbox{ GeV}\,,\quad
m_{\chi}\sim10^{-2}M\gtrsim10^{10}\mbox{ GeV}\,.
\ee
The splitting of the masses in the $\rho$ sector is much smaller than their central mass scale, so $m_{\rho}$ applies to both the fermionic and bosonic components of the $\rho$'s.  We find an important prediction from \eqref{eq:PVmessmass}: the existence of two pairs of new particles ($y$ and $\tilde{y}$) charged under the visible sector gauge group with masses independent of $M$ at about $1\mbox{ TeV}$.  These particles should be visible at the LHC as they would couple to the visible sector through Standard Model gauge interactions.

\begin{figure}[!th]
\begin{center}
\includegraphics[width=8cm]{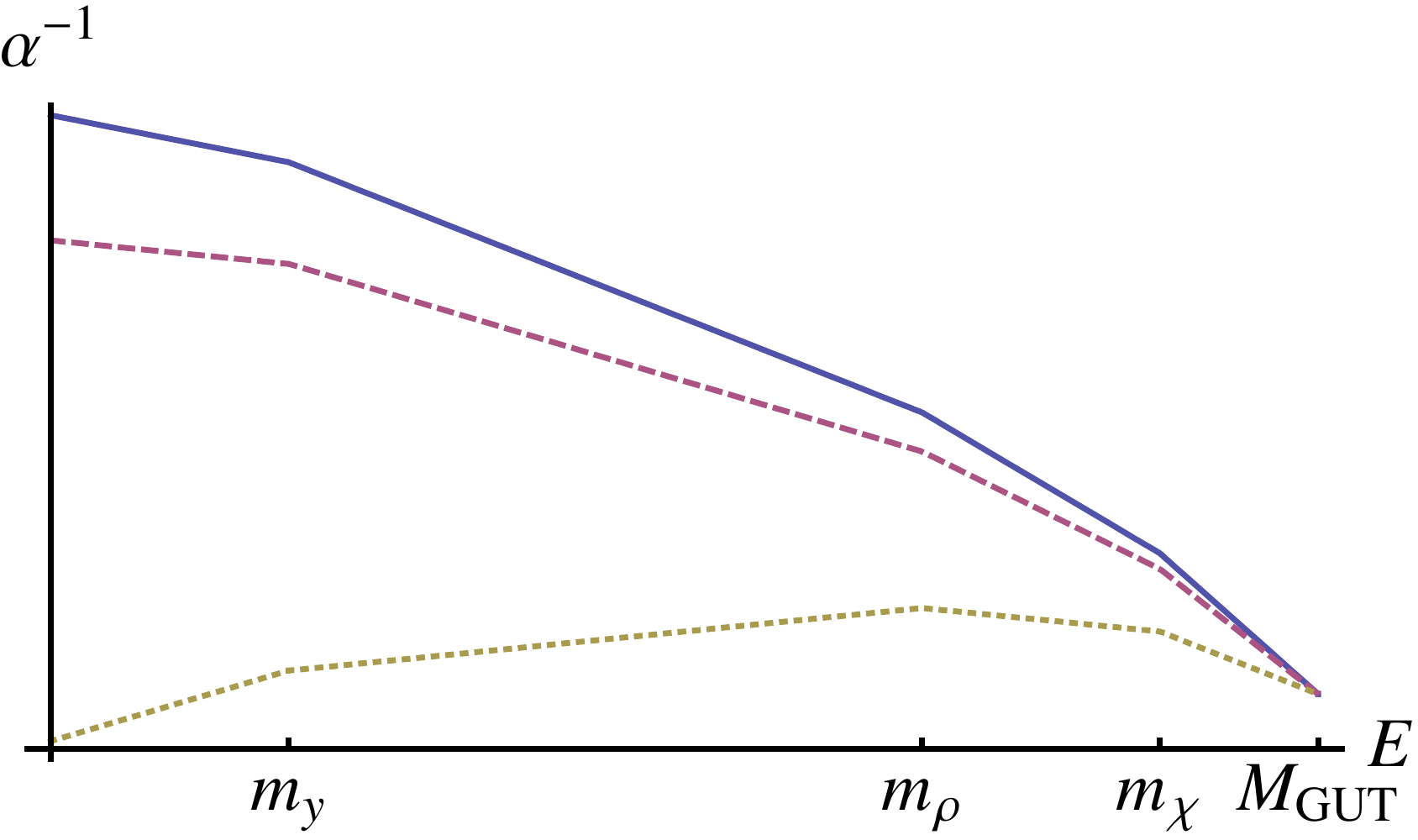}
\caption{\emph{The running of the gauge couplings in a typical $\SU{5}$ GUT.  The dotted line is for the strong force, the dashed for the weak and the solid for hypercharge.  As we reach various mass scales new particles from the SUSY breaking sector deflect the running.  As long as $M\gtrsim10^{16}\mbox{ GeV}$ we reach $M_{\rm GUT}\sim10^{16}\mbox{ GeV}$ before $\alpha_{\rm QCD}^{-1}$ reaches zero so avoid the Landau pole problem.}\label{fig:PVLP}}
\end{center}
\end{figure}

In the visible sector the first coupling constant to hit a Landau pole as we go up in energy in a standard $\SU{5}$ GUT will always be $\alpha_{\rm QCD}$ (see Figure \ref{fig:PVLP}).  The $\beta$-function coefficient for the QCD gauge coupling constant is
\be
b_{\rm QCD}=3-\Delta_{\rm mess}(E)
\ee
where $\Delta_{\rm mess}(E)$ is the contribution from the SUSY breaking sector as a function of the scale $E$.  It evaluates as
\be
\Delta_{\rm mess}(E)=\left\{\begin{array}{cl}
0 & E<m_y \\
2 & m_y\le E<m_{\rho} \\
4 & m_{\rho}\le E<m_{\chi} \\
7 & E \ge m_{\chi}\end{array}\right.
\ee
where the mass scales are defined as in \eqref{eq:PVmessmass}.  If we assume the QCD gauge coupling $\alpha_{\rm QCD}\sim10^{-1}$ at $m_y$ we find a Landau pole in the visible sector at
\begin{align}
& 10+\ln\nbrack{\frac{10^{-5}M}{10^3\mbox{ GeV}}}-\ln\nbrack{\frac{10^{-2}M}{10^{-5}M}}-4\ln\nbrack{\frac{E_{\rm LP}}{10^{-2}M}}=0\nonumber\\
\implies & E_{\rm LP}\sim10^{-4}\nbrack{\frac{M}{1\mbox{ GeV}}}^{5/4}\mbox{ GeV}\gtrsim10^{11}\mbox{ GeV}\,.
\end{align}
This is large enough to avoid any Landau pole problems up to a GUT scale of $M_{\rm GUT}\sim10^{16}\mbox{ GeV}$ for any high scale\footnote{The choice $M\sim10^{16}\mbox{ GeV}$ is actually quite interesting.  We would then have the Landau pole of the visible sector coinciding with both the Landau pole in the SUSY breaking sector and the GUT scale.  This would give no shortage of strong dynamical effects to generate the SUSY breaking terms.  Of course, reaching a Landau pole in the SUSY breaking sector suggests we should actually be working in the electric theory instead, so this analysis is not fully reliable.} $M\gtrsim10^{16}\mbox{ GeV}$.  For $M$ larger still the Landau pole is moved further and further above the GUT scale.  Note that when $M$ approaches the Planck mass this model exhibits the gauge-gravity hybrid behaviour discussed in \cite{Dudas:2008eq, Dudas:2008qf} (with the gravitational contribution possibly generating a $B$ parameter of order the weak scale).

\section{Conclusions\label{sec:conc}}

In this paper we have proposed an alternative method for stabilising the uplifted vacua of SQCD.  By restricting the magnetic gauge group to $\SU{2}$ we can stabilise all pseudomoduli with a baryon deformation to the superpotential.  The baryon deformation appears as a mass term in the magnetic superpotential so stabilises the pseudomoduli via the Coleman-Weinberg potential.  There remain tachyonic directions elsewhere in the pseudomoduli space so the model is able to produce gaugino masses comparable to sfermion masses when implicated in a direct mediation scenario.

The method is theoretically economical and all relevant quantities can be calculated and understood by simple, analytical expressions.  In addition, there are some strong phenomenological motivations for using this kind of model as a SUSY breaking sector.  The smallness of the gauge group and the emergent mass hierarchy in the SUSY breaking sector means there is no Landau pole problem.  The theory can be described by only two fundamental scales: the weak scale of $100\mbox{ GeV}$ and some high scale $M\gtrsim10^{16}\mbox{ GeV}$, which could easily be taken to be the GUT scale or the Planck scale and are understood through considerations of the UV theory.  The latter case approaches the gauge-gravity hybrid behaviour discussed in \cite{Dudas:2008eq, Dudas:2008qf}.  The deformations required to couple the SUSY breaking to the visible sector can be motivated by symmetry arguments if we presume the theory to possess an $R$-symmetry (which is broken explicitly by the same term responsible for SUSY breaking) and allow single trace operators of up to dimension six in the electric theory.  Furthermore, the model predicts an effective number of messengers of precisely $1/8$ and the existence of several new particles around $1\mbox{ TeV}$.  These particles would couple to the visible sector through Standard Model gauge interactions so could be seen at the LHC.

As well as the direct mediation scenario considered in depth here, other novel possibilities exist for this model.  The vacuum discussed in \S\ref{sec:BD} preserves an unbroken flavour group $\SU{2}\times\SU{N_f-2}$ which, for $N_f=5$, could be identified with the visible sector gauge group.

\vspace{5mm}\noindent{\bf Acknowledgements:}  I would like to thank S.\ Abel for discussions and helpful advice, and J.\ Jaeckel and V.V.\ Khoze for valuable comments.

\providecommand{\href}[2]{#2}\begingroup\raggedright\endgroup

\end{document}